\documentclass[
reprint,
superscriptaddress,
showpacs,preprintnumbers,
amsmath,amssymb,
aps,
]{revtex4-1}

\usepackage{color}
\usepackage{graphicx}
\usepackage{dcolumn}
\usepackage{bm}
\usepackage{txfonts}
\usepackage{tikz}
\usepackage{pgffor}
\usepackage{verbatim}
\usepackage{pstricks}
\usepackage{pst-3dplot}
\usepackage[colorlinks, breaklinks=true, linkcolor=blue, citecolor=blue, linktocpage=true]{hyperref}

\begin{document}

\title{
Voltage-driven Magnetization Switching via Dirac Magnetic Anisotropy and Spin--orbit Torque in Topological-insulator-based Magnetic Heterostructures
}
\date{\today}
\author{Takahiro Chiba}
 \affiliation{National Institute of Technology, Fukushima College, 30 Nagao, Kamiarakawa, Taira, Iwaki, 
Fukushima 970-8034, Japan}
\author{Takashi Komine}
 \affiliation{Graduate School of Science and Engineering, Ibaraki University, 4-12-1 Nakanarusawa, Hitachi, Ibaraki 316-8511, Japan}
 
\begin{abstract}
Electric-field control of magnetization dynamics is fundamentally and technologically important for future spintronic devices. Here, based on electric-field control of both magnetic anisotropy and spin--orbit torque, two distinct methods are presented for switching the magnetization in topological insulator (TI)/magnetic-TI hybrid systems. The magnetic anisotropy energy in magnetic TIs is formulated analytically as a function of the Fermi energy, and it is confirmed that the out-of-plane magnetization is always favored for the partially occupied surface band. Also proposed is a transistor-like device with the functionality of a nonvolatile magnetic memory that uses voltage-driven writing and the (quantum) anomalous Hall effect for readout. For the magnetization reversal, by using parameters of Cr-doped ${\rm (Bi_{1-x}Sb_{x})_{2}Te_{3}}$, the estimated source-drain current density and gate voltage are of the orders of $10^4$--$10^5$~A/cm$^2$ and 0.1~V, respectively, below 20~K and the writing requires no external magnetic field. Also discussed is the possibility of magnetization switching by the proposed method in TI/ferromagnetic-insulator bilayers with the magnetic proximity effect.
\end{abstract}


\maketitle

\section{Introduction}

Electrical control of magnetism is essential for the next generation of spintronic technologies, such as nonvolatile magnetic memory, high-speed logic, and low-power data transmission \cite{Hoffmann15}. In these technologies or devices, the magnetization direction of a nanomagnet is controlled by an electrically driven {\it torque} rather than an external magnetic field. A representative torque is the current-induced spin--orbit torque (SOT) \cite{Manchon19} in heavy-metal/ferromagnet heterostructures, wherein the spin Hall effect in the heavy metal \cite{Liu12, Kim12} and/or the Rashba--Edelstein effect (also known as the inverse spin-galvanic effect) at the interface \cite{Miron11} play crucial roles in generating the torque. Recently, several experiments have reported a giant SOT efficiency in topological insulator (TI)-based magnetic heterostructures such as both TI/magnetic-TI \cite{Fan14} and TI/ferromagnetic-metal hybrid systems \cite{Mellnik14, HYang15}. A TI has a metallic surface state in which the spin and momentum are strongly correlated (known as spin--momentum locking) because of a strong spin--orbit interaction in the bulk state \cite{Pesin12, Ando13}, which is expected to lead to the giant SOT \cite{Kondou16}. Indeed, magnetization reversal by SOT has been proposed theoretically \cite{Garate10, Yokoyama11, Mahfouzi16, Ndiaye17} and demonstrated experimentally in magnetic TIs \cite{Fan14, Fan16, Yasuda17} as well as TI/ferromagnet bilayers \cite{Han17, Wang17, Dc18, Wu19A, Wu19, Li19S}. Remarkably, the critical current density required for switching is of the order of $10^5$~A/cm$^2$, which is much smaller than the corresponding values ($10^6$--$10^8$~A/cm$^2$) for heavy-metal/ferromagnet heterostructures \cite{Liu12, Kim12, Miron11}. In particular, the magnetization switching of magnetic TIs is more efficient: {\it Yasuda}~{\it et~al.}\ \cite{Yasuda17} succeeded in reducing the switching current density by means of a current pulse injected parallel to a bias magnetic field, whereas {\it Fan}~{\it et~al.}\ \cite{Fan16} realized magnetization reversal by means of a scanning gate voltage with a small constant current and in-plane magnetic field.

Another important method for controlling magnetic properties is the electric-field effect in magnets, such as controlling ferromagnetism in dilute magnetic semiconductors \cite{Chiba08}, manipulating magnetic moments in multiferroic materials \cite{Chu08}, and changing the magnetic anisotropy in an ultrathin film of ferromagnetic metal \cite{Weisheit07, Duan08, Maruyama09}. In particular, voltage control of magnetic anisotropy (VCMA) in ferromagnets promises energy-efficient reversal of magnetization by means of what is known as voltage torque, which has been demonstrated by using a pulsed voltage under a constant-bias magnetic field in a magnetic tunnel junction \cite{Shiota12}. This approach is based on the clocking scheme in which one first sets the ferromagnet to an initial stable state under the application of an external bias and then inputs the signal voltage pulse to determine the final state. Recent experiments using heavy-metal/ferromagnet/oxide heterostructures have demonstrated that the critical current for SOT-driven switching of perpendicular magnetization can be modulated by an electric field via VCMA \cite{Baek18, Mishra19}. By contrast, the electric-field effect in a magnetic TI \cite{Wang15, Sekine16} and a TI/ferromagnetic-insulator (FI) bilayer \cite{Semenov12, Flatte17} has been investigated to date in terms of the voltage-torque-driven magnetization dynamics. Note that {\it Semenov}~{\it et~al.}\ \cite{Semenov12} demonstrated magnetization rotation between the in-plane and out-of-plane directions by VCMA at the TI/FI interface. Therefore, it becomes highly desirable to control the magnetic anisotropy and SOT simultaneously by means of the electric field in TI-based magnetic heterostructures, which may lead to magnetization switching that is more energetically efficient.

In this paper, inspired by the SOT and VCMA approaches for magnetization control, we combine them and present two distinct clocking methods for magnetization switching in TI/magnetic-TI hybrid systems. First, we model the current-induced SOT and magnetic anisotropy energy (MAE) in TI-based magnetic heterostructures as a function of the Fermi energy to determine a stable magnetization direction at the electrostatic equilibrium. Then we propose a transistor-like device with the functionality of a nonvolatile magnetic memory that uses (i) VCMA writing that requires no external magnetic field and (ii) readout based on the anomalous Hall effect. For the magnetization reversal, we estimate the source-drain current density and gate voltage. Finally, we show the switching phase diagram for the input pulse width and voltages as a guide to realizing the proposed method of magnetization reversal. We also discuss the possibility of using the proposed method for magnetization reversal in TI/FI bilayers with magnetic proximity \cite{Jiang15, Hirahara17, Fanchiang18} at the interface.
 
\section{Model}\label{model}

We begin this section by deriving the current-induced SOT in TI-based magnetic heterostructures by using the current-spin correspondence of two-dimensional (2D) Dirac electrons on the TI surface. Next, in the same system we formulate the MAE analytically to determine a stable magnetization direction at the electrostatic equilibrium and to reveal the controllability of the VCMA effect. To model the SOT and VCMA, we consider 2D massless Dirac electrons on the TI surface, which is exchange coupled to the homogeneous localized moment of a magnetic TI (or an attached FI as discussed in Sec.~\ref{TI/FI}). When the surface electrons interact with the localized moment, they have an exchange interaction that can be modeled by a constant spin splitting $\Delta$ along the magnetization direction with unit vector ${\bf m} = {\bf M}/M_{\rm s}$ (in which ${\bf M}$ is the magnetization vector with the saturation magnetization $M_{\rm s}$) \cite{Nomura10}. Then, the following 2D Dirac Hamiltonian provides a simple model for the electronic structure of the TI surface state:
\begin{align}
\mathcal{H}_{\bf k} = \hbar v_{\rm F}\hat{\bm \sigma}\cdot\left({\bf k}\times\hat{\bf z}\right)+\Delta\hat{\bm \sigma}\cdot{\bf m}, \label{H2d}
\end{align}
where $h = 2\pi\hbar$ is the Planck constant, $v_{\rm F}$ is the Fermi velocity of the Dirac electrons, $\hat{\bm \sigma}$ is the Pauli matrix operator for the spin, and $\Delta$ is the exchange interaction. For simplicity, we ignore here the particle--hole asymmetry in the surface bands. Introducing the polar angle $\theta$ and azimuthal angle $\varphi$ for ${\bf m}=(\cos\varphi\sin\theta, \sin\varphi\sin\theta, \cos\theta)$, the energy dispersion of the Hamiltonian (\ref{H2d}) can be expressed as
\begin{align}
E_{{\bf k}s} = s\sqrt{\left( \hbar v_{\rm F}k\right)^2 + \Delta^2 -2\hbar v_{\rm F}k\Delta\sin\theta\sin\left( \varphi - \varphi_k\right)}, \label{E2d}
\end{align}
where $s=\pm$ corresponds to the upper and lower bands, and $\cos\theta_{\bf k}=\Delta\cos\theta/|E_{{\bf k}s}|$ and $\tan\varphi_{\bf k}=k_y/k_x$ are the polar and azimuthal angles of the spinors on the Bloch sphere, respectively.

\subsection{Current--induced spin--orbit torque}\label{SOT}

\begin{figure}[ptb]
\begin{centering}
\includegraphics[width=0.45\textwidth,angle=0]{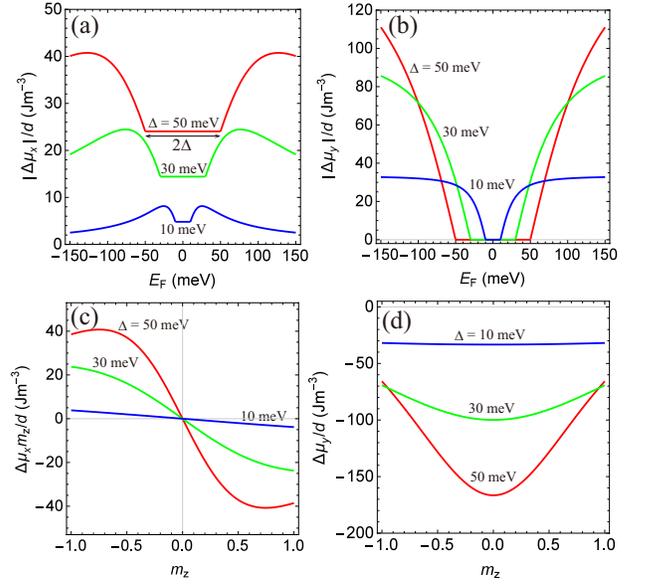} 
\par\end{centering}
\caption{
Current-induced nonequilibrium spin polarization $\Delta{\bm \mu}/d$ scaled by thickness $d$ of ferromagnet as a function of $E_{\rm F}$ for different values of $\Delta$~(with $m_z = 1$): (a) $x$-component $\Delta\mu_x/d$; (b) $y$-component $\Delta\mu_y/d$; (c)~$\Delta\mu_x/d$ and (d)~$\Delta\mu_y/d$ at $E_{\rm F} = 95$~meV (corresponding carrier density $\sim10^{12}$~cm$^{-2}$) as functions of $m_z$ for different values of $\Delta$. 
In these graphs, we use $v_{\rm F} = 4.0\times10^{5}$~ms$^{-1}$, $d = 10$~nm, and $E_x = 0.1$~V/$\mu$m. The details of the calculations are given in the text.
}
\label{fig:spin}
\end{figure}

We begin by discussing a current-induced SOT to the magnetization in TI-based magnetic heterostructures \cite{Garate10, Yokoyama10, Sakai14, Mahfouzi16, Ndiaye17, Chiba17, Ghosh18}. The SOT stems from the exchange interaction between the magnetization and the electrically induced nonequilibrium spin polarization ${\bm \mu}$ (in units of m$^{-2}$) \cite{Mellnik14, Kondou16}, which can be described by
\begin{align}
{\bf T}_{\rm SO} = -\gamma\frac{\Delta{\bm \mu}}{M_{\rm s}d}\times{\bf m}, \label{SOT}
\end{align}
where $\gamma$ is the gyromagnetic ratio and $d$ is the thickness of the ferromagnetic layer (magnetic TI). In short, the SOT is obtained by calculating the electrically induced spin polarization on the TI surface.

As a characteristic feature of the Dirac Hamiltonian (\ref{H2d}), the spin operator $\hat{\bm \sigma}$ is directly proportional to the velocity operator $\hat{\bf v} = \partial\mathcal{H}_{\bf k}/(\hbar\partial{\bf k}) = v_{\rm F}\hat{\bf z}\times\hat{\bm \sigma}$ due to the spin-momentum lock. In this sense, we can identify the nonequilibrium spin polarization ${\bm \mu}$ with the electric current ${\bf J}$ on the TI surface, namely 
\begin{align}
{\bm \mu} = -\frac{1}{ev_{\rm F}}\hat{\bf z}\times{\bf J}, \label{spin}
\end{align}
where $-e~(e>0)$ is the electron charge. In the following, we use ${\bm \mu}$ and ${\bf J}$ to denote the quantum statistical expectation values of $\hat{\bm \sigma}$ and $\hat{\bf j} = -e\hat{\bf v}$, respectively. Here, we emphasize that the nonequilibrium spin polarization involves only in-plane spin components. Hence, the in-plane component of the spin susceptibility corresponds to the electric conductivity via Eq.~(\ref{spin}). In the framework of Boltzmann transport theory and the Kubo formula, previous studies \cite{Sakai14, Chiba17, Sinitsyn07, Culcer11, Ado15, Sabzalipour15} have calculated the longitudinal and transverse (anomalous Hall) conductivities on magnetized TI surfaces by assuming a short-range impurity potential with Gaussian correlations $\langle\hat{V}({\rm r}_{1})\hat{V}({\rm r}_2)\rangle_{\rm imp} = nV_0^2\delta({\rm r}_1-{\rm r}_2)$ in which $n$ is the impurity concentration, $V_0$ is the scattering potential, and $\langle\cdots\rangle_{\mathrm{imp}}$ indicates an ensemble average over randomly distributed impurities. For an electric field ${\bf E}$ along the TI surface, the driving sheet current can be written as ${\bf J} = \sigma_{\rm L}{\bf E} + \sigma_{\rm AH}\hat{\bf z}\times{\bf E}$ with \cite{Chiba17} 
\begin{align}
\begin{split}
\sigma_{\rm L} 
& = \frac{e^2}{2h}\frac{E_{\rm F}\tau}{\hbar}\frac{1-\xi^{2}m_{z}^2}{1+3\xi^2m_{z}^2}, \\
\sigma_{\rm AH} 
& = -\frac{4e^2}{h}\xi m_{z}\frac{1+\xi^2m_{z}^2}{\left( 1+3\xi^2m_{z}^2\right)^2}, \label{sigma}
\end{split}
\end{align}
where $\tau = 4\hbar\left( \hbar v_{\rm F}\right)^2/\left( nV_0^2E_{\rm F}\right)$ is the transport relaxation time of massless Dirac electrons within the Born approximation, $\xi = \Delta/E_{\rm F}$ with the Fermi energy $E_{\rm F}$ measured from the original band-teaching (Dirac) point, and $m_z$ denotes the $z$-component of ${\bf m}$. Note that $\sigma_{\rm AH}$ in Eq.~(\ref{sigma}) is independent of the impurity parameters but diagrammatically contains the side-jump and skew scattering contributions as well as the intrinsic one associated with the Berry curvature of the surface bands \cite{Sinitsyn07, Ado15}. According to Eq.~(\ref{spin}), the current-induced spin polarization for ${\bf E} = E_x\hat{\bf x}$ is therefore 
\begin{align}
{\bm \mu} = -\frac{1}{ev_{\rm F}}\left( -\sigma_{\rm AH}{\bf E} + \sigma_{\rm L}\hat{\bf z}\times{\bf E}\right) \equiv \mu_xm_z\hat{\bf x} + \mu_y\hat{\bf y}, \label{spin2}
\end{align}
where 
\begin{align}
\mu_x 
& = -\frac{4eE_x}{hv_{\rm F}}\Delta\frac{E_{\rm F}\left( E_{\rm F}^2+\Delta^2m_{z}^2\right)}{\left( E_{\rm F}^2+3\Delta^2m_{z}^2\right)^2}, \label{mux}\\
\mu_y 
& = -\frac{eE_x}{2hv_{\rm F}}\frac{E_{\rm F}\tau}{\hbar}\frac{E_{\rm F}^2-\Delta^{2}m_{z}^2}{E_{\rm F}^2+3\Delta^2m_{z}^2}.\label{muy}
\end{align}
Equations~(\ref{mux}) and (\ref{muy}) are substantially equivalent to the current-induced nonequilibrium spin density that {\it Ndiaye}~{\it et~al.}\ calculated directly by using the Kubo--Streda formula involving the spin vertex correction \cite{Ndiaye17}. Note that $\mu_x$ originates from the magnetoelectric coupling (the so-called Chern--Simons term) \cite{Nomura10, Garate10} that is proportional to the anomalous Hall conductivity [see Eq.~(\ref{spin2})]. Meanwhile, $\mu_y$ stems from the Rashba--Edelstein effect due to the spin-momentum locking on the TI surface \cite{Yokoyama10}.

From Eqs.~(\ref{SOT}) and (\ref{spin2}), we finally obtain the form of SOT arising from the TI surface \cite{Ndiaye17, Chiba17} [see the Appendix for the current expression of SOT], namely 
\begin{align}
{\bf T}_{\rm SO} = \gamma\frac{\Delta \mu_x}{M_{\rm s}d}m_z{\bf m}\times\hat{\bf x} + \gamma\frac{\Delta \mu_y}{M_{\rm s}d}{\bf m}\times\hat{\bf y}.\label{SOT2}
\end{align}
The first term contributes as a damping-like (DL) torque but one that is quite different from that of the spin Hall effect in traditional heavy-metal/ferromagnet heterostructures \cite{Liu12, Kim12}. In fact, for the in-plane magnetization configuration ($m_z = 0$), this DL torque vanishes because of the absence of the magnetoelectric coupling via the anomalous Hall effect, whereas the SOT driven by the spin Hall effect acts on the magnetization. Meanwhile, despite its origin, the second term acts as only a field-like (FL) torque. This feature is also different from that of the Rashba--Edelstein effect in the usual 2D ferromagnetic Rashba systems in which there might be both FL and DL contributions \cite{Gao15, Lee15}. Figure~\ref{fig:spin}(a) and (b) show the $E_{\rm F}$ dependence of the $x$- and $y$-components, respectively, of $\Delta{\bm \mu}$ (in units of Jm$^{-2}$) for different values of the surface band gap. For this calculation, $\Delta$ is used within the values reported experimentally in magnetically doped \cite{Tokura19} and FI-attached \cite{Hirahara17, Mogi19} TIs. We also adopt $n = 10^{12}$~cm$^{-2}$ and $V_0 = 0.2$~keV\AA$^2$ as impurity parameters based on an analysis of the transport properties of a TI surface \cite{Chiba19}. These impurity parameters can reproduce the experimentally observed longitudinal resistance ($\sim10$~k$\Omega$) in magnetic TIs. Remarkably, as seen in Fig.~\ref{fig:spin}(a), even when the Fermi level is inside the surface band gap ($E_{\rm F} < |\Delta|$), the $x$-component survives as \cite{Garate10, Sakai14}
\begin{align}
\mu_x 
= -\frac{eE_x}{2hv_{\rm F}} = -\frac{1}{ev_{\rm F}}\sigma_{\rm QAH}E_x, \label{muxgap}
\end{align}
where $\sigma_{\rm QAH} = e^2/(2h){\rm sgn}(m_z)$ characterizes the quantum anomalous Hall effect on the magnetized TI surface \cite{Chang13S}, reflecting the topological nature of 2D massive Dirac electrons. By contrast, because of the Rashba--Edelstein effect, the $y$-component shown in Fig.~\ref{fig:spin}(b) survives in only the metallic surface states ($E_{\rm F} \geq |\Delta|$). Figure~\ref{fig:spin}(c) and (d) show the $m_z$ dependence of the $x$- and $y$-components, respectively, of $\Delta{\bm \mu}$ for different values of the surface band gap. In these plots, we include $m_z$ in the $x$-component of $\Delta{\bm \mu}$. Reflecting the anomalous Hall effect on the magnetized TI surface, the $x$-component is odd upon magnetization reversal, whereas the $y$-component is even in magnetization reversal because it is proportional to $\sigma_{\rm L}$ via the Rashba--Edelstein effect.

\subsection{Dirac magnetic anisotropy}\label{MA}

Here, to evaluate the VCMA effect in TI-based magnetic heterostructures, we investigate the MAE associated with the exchange interaction in Eq.~(\ref{H2d}). The MAE is defined as the difference in the sums over occupied states of energy dispersions (\ref{E2d}) with $\theta=0$ as the reference state \cite{Ieda18}, namely 
\begin{align}
U_{\rm MAE} = \sum_{{\bf k}s}^{\rm occ.}E_{{\bf k}s}(\theta) - \sum_{{\bf k}s}^{\rm occ.}E_{{\bf k}s}(\theta=0).\label{UMAE}
\end{align}
Expanding Eq.~(\ref{UMAE}) around $\theta \approx 0$ leads to $U_{\rm MAE} \approx K_u\sin^2\theta$, where the uniaxial magnetic anisotropy constant $K_u$ (in units of Jm$^{-2}$) is given by 
\begin{align}
K_u = -\sum_{{\bf k}s}^{\rm occ.}s\frac{\left( \hbar v_{\rm F}k\right)^2\Delta^2\sin^2\left( \varphi - \varphi_{{\bf k}}\right)}{2\left[ \left( \hbar v_{\rm F}k\right)^2 + \Delta^2\right]^{3/2}}.\label{Ku}
\end{align}
The sign of $K_u$ specifies the type of MAE, namely perpendicular magnetic anisotropy (PMA, $K_u>0$) or easy-plane magnetic anisotropy ($K_u<0$). For the partially occupied energy bands, we have $K_u>0$; i.e., PMA is always favored by the magnetization coupled with Dirac electrons on the TI surface. A qualitative understanding of the characteristic PMA is given by a gain of electronic free energy associated with the exchange interaction between the Dirac electrons and localized moment. When the magnetization is along the out-of-plane direction, a surface band gap ($2\Delta m_z$) emerges in the massless Dirac dispersion, which reduces the electron group velocity (kinetic energy). Meanwhile, for the in-plane magnetization orientation, the exchange interaction merely shifts the surface band in the $k$-space. In terms of the exchange interaction maximizing the energy gain of the Dirac electron system, the case possessing the surface band gap is expected to be more favorable with lower electronic free energy than that with the shifted surface bands by an in-plane magnetization, whose scenario can be interpreted as being analogous to the Peierls transition in electron--lattice coupled systems \cite{Tokura19}.

\begin{figure}[ptb]
\begin{centering}
\includegraphics[width=0.45\textwidth,angle=0]{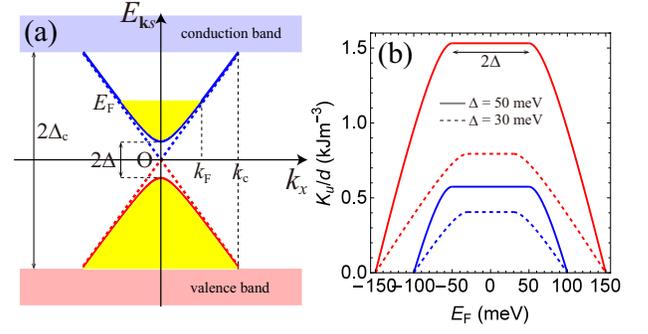} 
\par\end{centering}
\caption{
(a) Schematic of massless (dashed line) and massive (solid line) surface state dispersions at $k_y = 0$ in which $E_{\rm F}$ denotes the Fermi energy measured from the Dirac point ($E_{{\bf k}s}=0$) of the original massless surface bands, $2\Delta$ is the surface band gap due to an exchange interaction, and $2\Delta_{\rm c}$ is the bulk band gap. $k_{\rm F}$ and $k_{\rm c}$ correspond to the Fermi wave vector and cutoff wave vector, respectively. (b) Scaled magnetic anisotropy energy $K_u/d$ as a function of $E_{\rm F}$ for different values of $\Delta_{\rm c}$ and $\Delta$. Red and blue lines are for $\Delta_{\rm c} = 150$~meV and $\Delta_{\rm c} = 100$~meV, respectively. In this plot, we use $v_{\rm F} = 4.0\times10^{5}$~ms$^{-1}$ and $d = 10$~nm.
}
\label{fig:band Ku}
\end{figure}

To integrate Eq.~(\ref{Ku}), we assume hereinafter that the low-energy Dirac Hamiltonian (\ref{H2d}) is a valid description for $k \leq k_{\rm c}$ with a momentum cut $k_{\rm c} = \sqrt{\Delta_{\rm c}^2 - \Delta^2}/(\hbar v_{\rm F})$ \cite{Tserkovnyak15} in which $2\Delta_{\rm c}$ is the bulk band gap of TIs induced by the band inversion due to the spin--orbit interaction. We also define the Fermi wave vector $k_{\rm F} = \sqrt{E_{\rm F}^2 - \Delta^2}/(\hbar v_{\rm F})$, as shown in Fig.~\ref{fig:band Ku}(a). Without loss of generality, we assume the case in which $E_{\rm F}$ crosses the upper surface band, namely $E_{\rm F} > \Delta$. Then, we obtain the magnetic anisotropy constant by integrating over the relevant energy range as
\begin{align}
K_u
& = \frac{\Delta^2}{8\pi(\hbar v_{\rm F})^2}\left[ \Delta_{\rm c} - E_{\rm F} - \left( \frac{1}{E_{\rm F}} - \frac{1}{\Delta_{\rm c}}\right)\Delta^2\right].\label{Kua}
\end{align}
Up to the lowest order of $\Delta$, Eq.~(\ref{Kua}) takes the simplest analytic from of $K_u = \Delta^2k_{\rm c}/(8\pi\hbar v_{\rm F})$, which corresponds to the out-of-plane MAE for $E_{\rm F} \leq |\Delta|$ derived earlier by {\it Tserkovnyak}~{\it et~al.}\ \cite{Tserkovnyak15}. Figure~\ref{fig:band Ku}(b) shows the $E_{\rm F}$ dependence of $K_u$ for different values of the bulk and surface band gaps. In this plot, $\Delta_{\rm c} = 150$~meV and $\Delta_{\rm c} = 100$~meV correspond to the bulk band gaps for ${\rm Bi_{2-x}Sb_{x}Te_{3-y}Se_{y}}$ (BSTS) and ${\rm (Bi_{1-x}Sb_{x})_{2}Te_{3}}$ (BST) \cite{Ando13}, respectively. Because we assume that the surface states have energy dispersions with particle--hole symmetry, the magnetic anisotropy constant retains the form of Eq.~(\ref{Kua}) in the case in which $E_{\rm F}$ crosses the lower surface band, namely $E_{\rm F} < -\Delta$. Hence, $K_u$ is maximum with the form of $K_u|_{E_{\rm F} = |\Delta|}$ for $E_{\rm F} \leq |\Delta|$ and decreases apart from the energy level of $\pm\Delta$. The reason is that there are simply fewer active electrons for $E_{\rm F} < -\Delta$, while for $E_{\rm F} > \Delta$ the energy decrease in the lower band is compensated partially by the upper-band energy increase in conductive electrons.

At the end of this section, we compare the formulated MAE with a recent experiment employing Cr-doped BST thin films which is sandwiched by two different dielectrics and hence has the Dirac electron systems on each interface \cite{Fan16}. 
The magnetic anisotropy field due to Eq.~(\ref{Kua}) is defined by $B_K = 2K_u/(M_{\rm s}d)$. For the Cr-doped BST films with 7 quintuple-layer ($d \approx 7$~nm), a calculated result is $B_K = 117$~mT~($K_u/d = 0.5$~kJm$^{-3}$) while an experiment reports $B_K \approx 570$~mT without an electric gate in Ref.~\onlinecite{Fan16}. In this calculation, we choose the parameters for Cr-doped BST: $\Delta_{\rm c} = 100$~meV, $\Delta = 30$~meV, $v_{\rm F} = 4.0\times10^{5}$~ms$^{-1}$, $M_{\rm s} = 8.5\times10^3$~Am$^{-1}$, and an electron/hole carrier density $1.0/0.2\times10^{12}$~cm$^{-2}$~(corresponding to $|E_{\rm F}| = 98/51$~meV) for the each interface \cite{Fan16}. The difference of $B_K$ might come from disregarding the realistic particle--hole asymmetry induced by the higher--order $k$-term of the energy dispersion which makes a more sharp electronic density of states in the surface valence band and enhances the hole--mediated Dirac PMA up to $K_u/d = 8$~kJm$^{-3}$ \cite{Semenov12}. In this respect, our simple Dirac model could not reflect the detail of the realistic surface band but could capture the permissible magnitude of $B_K$. 
Therefore, the above comparison implies that the interfacial Dirac PMA gives a significant contribution to the magnetic anisotropy in dilute magnetic TIs.

\begin{figure}[ptb]
\begin{centering}
\includegraphics[width=0.45\textwidth,angle=0]{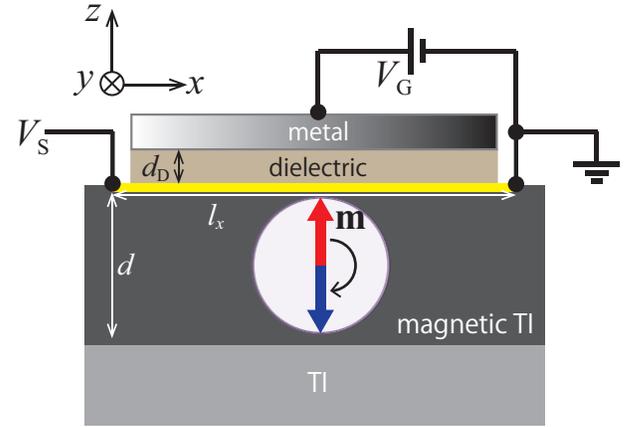} 
\par\end{centering}
\caption{
Schematic geometry (side view) of field-effect transistor (FET)-like device comprising a magnetic topological insulator (TI) film (with thickness $d$) sandwiched by a nonmagnetic TI and a dielectric attached to a top electric gate $V_{\rm G}$ in which the Dirac electron system should appear on the top surface of the magnetic TI \cite{Yasuda17}. $d_{\rm D}$ is the thickness of the dielectric and $l_x$ is the length of the conduction channel. $V_{\rm S}$ is the voltage difference between the source and drain electrodes. Current flows on the $x$--$y$ plane depicted by a yellow line that corresponds to the TI surface state. The arrows denote the initial (red) and final (blue) magnetization directions in the magnetization reversal.
}
\label{fig:device}
\end{figure}

\begin{figure*}[ptb]
\begin{centering}
\includegraphics[width=0.99\textwidth,angle=0]{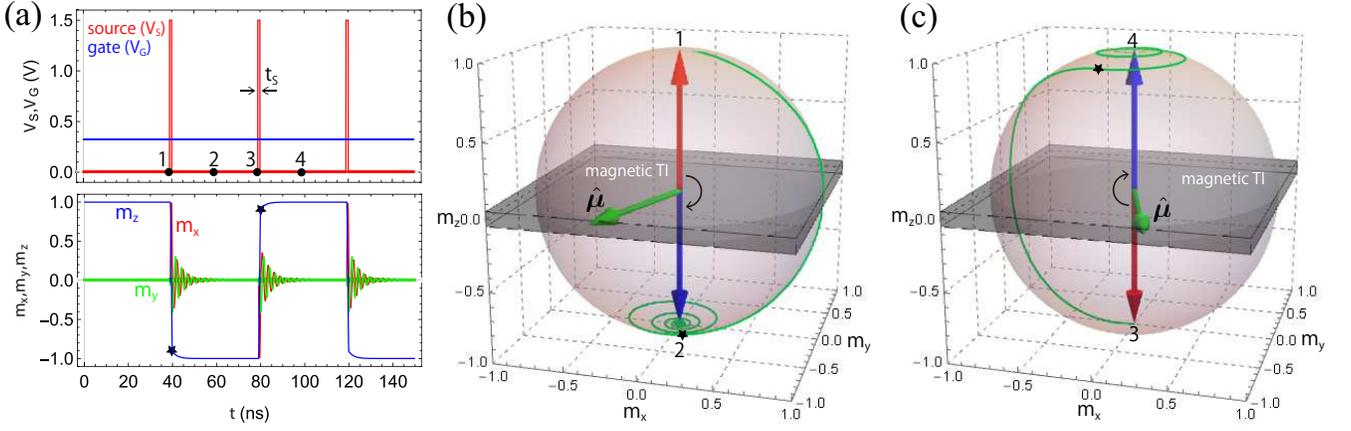} 
\par\end{centering}
\caption{
(a) Time evolution of each component of magnetization with pulsed source voltage ($V_{\rm S}$). The numerical calculation is performed with static $V_{\rm G} = 0.32$~V, $V_{\rm S} = 1.5$~V, and $t_{\rm S} = 1$~ns. (b), (c) Corresponding magnetization switching trajectories during duration of (b) $1 \to 2$ and (c) $3 \to 4$ in left upper panel. The vertical arrows denote the initial (red) and final (blue) magnetization directions in the magnetization reversal. The green arrow on the surface of the magnetic TI (black cube) indicates the direction of the current-induced spin polarization: $\hat{\bm \mu} = {\bm \mu}/|{\bm \mu}|$. The spin--orbit torque (SOT) is active from 1(3) to the black star ($\star$).
}
\label{fig:VS}
\end{figure*}

\section{Magnetization switching}\label{switching}

To demonstrate magnetization switching via SOT and VCMA, we propose a field-effect transistor (FET)-like device with a magnetic TI film as a conduction channel layer \cite{Fan16} in which source-drain ($V_{\rm S}$) and gate ($V_{\rm G}$) voltages are applied, as shown in Fig.~\ref{fig:device}. To investigate the macroscopic dynamics of the magnetization in the device, we solve the Landau--Lifshitz--Gilbert (LLG) equation including the SOT (\ref{SOT2}), namely 
\begin{align}
\frac{d\bf m}{dt} = -\gamma{\bf m}\times
{\bf B_{\rm eff}} + \alpha_{\rm eff}{\bf m}\times\frac{d\bf m}{dt} + {\bf T}_{\rm SO}(V_{\rm G})
, \label{LLG}
\end{align}
where ${\bf B_{\rm eff}}$ is an effective magnetic field obtained by finite ${\bf m}$ functional derivatives of the total energy $U_{\rm M}$, namely ${\bf B_{\rm eff}} = -\delta U_{\rm M}/(M_{\rm s}\delta{\bf m})$, and $\alpha_{\rm eff}$ denotes the effective Gilbert damping constant. 
As discussed in Sec.~\ref{MA}, the interfacial Dirac PMA gives a significant contribution to the magnetic anisotropy in in dilute magnetic TIs. Because of the thin magnetic TI in Fig.~\ref{fig:device}, we assume that $U_{\rm M}$ consists of the MAE (\ref{Kua}) and the magnetostatic energy that generates a demagnetization field, i.e., 
\begin{align}
U_{\rm M} = \frac{1}{d}K_u(V_{\rm G})\left( 1 - m_z^2\right) + \frac{1}{2}\mu_0M_{\rm s}^2m_z^2, \label{EM}
\end{align}
where $\mu_0$ is the permeability of free space. As discussed in Sec.~\ref{model}, both MAE and SOT depend on the position of $E_{\rm F}$, which can be controlled electrically via (see the Appendix for details) 
\begin{align}
E_{\rm F}(V_{\rm G}) = \hbar v_{\rm F}\sqrt{4\pi\left( n_{\rm int} + \frac{\Delta^2}{4\pi\left( \hbar v_{\rm F}\right)^2} + \frac{\epsilon}{ed_{\rm D}}V_{\rm G}\right)}, \label{EVG} 
\end{align}
where $\epsilon$ is the permittivity of a dielectric of thickness $d_{\rm D}$ and $n_{\rm int}$ is the intrinsic carrier density at $V_{\rm G} = 0$. In this study, for the dielectric layer with $d_{\rm D} = 20$~nm, we adopt a typical insulator, namely Al$_2$O$_3$ (for which the relative permittivity is $\epsilon/\epsilon_0 = 9.7$) \cite{Yoshimi15, Fan16}.

Because the ferromagnetic Curie temperature ($T_{\rm c}$) of Cr-doped BST is less than 35~K \cite{Chang13}, we first consider magnetization switching at zero temperature. Influence of finite temperatures on the magnetization switching will be discussed in Sec.~\ref{temperature}.
The equilibrium magnetization direction with neither $V_{\rm S}$ nor $V_{\rm G}$ is determined by minimizing Eq.~(\ref{EM}) regarding the polar angle $\theta$. However, for simplicity, we assume that $\theta \approx 0^\circ$ at the electrostatic equilibrium. This assumption is permissible because hereinafter we consider a magnetic TI such as Cr-doped BST with a very small $M_{\rm s} = 8.5\times10^3$~Am$^{-1}$ \cite{Fan16}, neglecting the demagnetizing field effect from the second term in Eq.~(\ref{EM}) at $V_{\rm G} = 0$. 
For numerical simulation, Eq.~(\ref{LLG}) is solved by setting $\theta = 1^\circ$ and $\varphi = 0^\circ$ as the initial condition for ${\bf m}(t=0)$. In the simulation, we choose the parameters for Cr-doped BST as $\Delta_{\rm c} = 100$~meV, $\Delta = 30$~meV, $v_{\rm F} = 4.0\times10^{5}$~ms$^{-1}$, $d = 7$~nm, $n = 10^{12}$~cm$^{-2}$, $V_0 = 0.2$~keV\AA$^2$ \cite{Chiba19}, $n_{\rm int} \approx 0$~cm$^{-2}$ (corresponding to $E_{\rm F}(V_{\rm G} = 0) = \Delta = 30$~meV) \cite{Yoshimi15,Chang13}, $\gamma = 1.76\times10^{11}$~T$^{-1}$s$^{-1}$, and $\alpha_{\rm eff} = 0.1$. Note that a large enhanced damping from 0.03 to 0.12 due to the strong spin--orbit interaction of TIs has been reported in TI/ferromagnetic-metal bilayers \cite{Jamali15}. In addition, no modulation of $\alpha_{\rm eff}$ is assumed during the duration of $V_{\rm G}$ because we consider a thicker magnetic TI film ($d = 7$~nm) than the ferromagnetic metal used in the magnetic tunnel junction \cite{Okada14}.

\begin{figure*}[ptb]
\begin{centering}
\includegraphics[width=0.99\textwidth,angle=0]{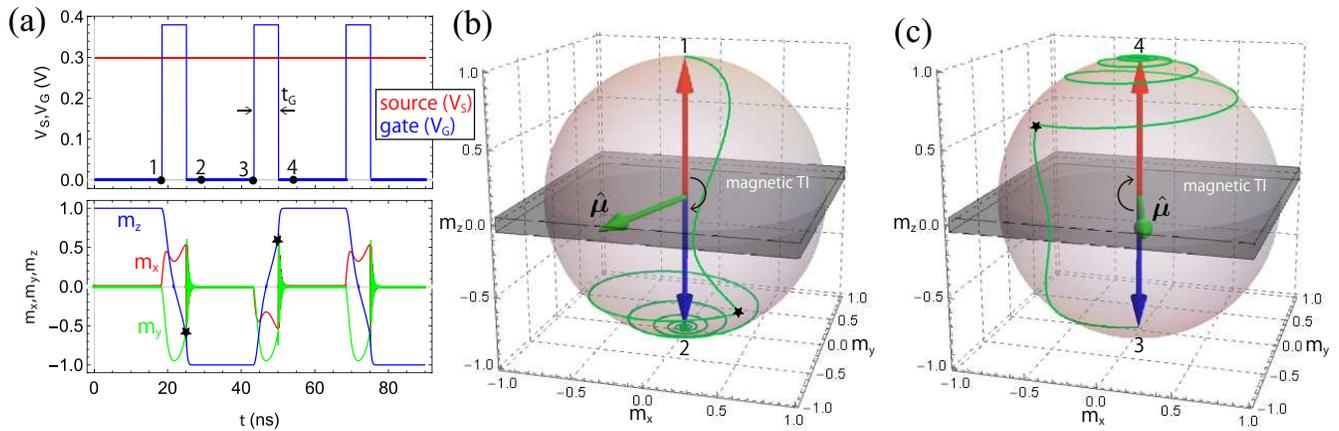} 
\par\end{centering}
\caption{
(a) Time evolution of each component of magnetization with pulsed gate voltage $V_{\rm G}$. The numerical calculation is performed with $V_{\rm G} = 0.38$~V, $V_{\rm S} = 0.3$~V, and $t_{\rm G} = 6.7$~ns. (b), (c) Corresponding magnetization switching trajectories during (b) $1 \to 2$ and (c) $2 \to 3$ in the left upper panel. The red, blue, and green arrows and the black star have the same meanings as in Fig.~\ref{fig:VS}.
}
\label{fig:VG}
\end{figure*}

\subsection{Switching via a source-drain current pulse $J_{\rm S}$}\label{S pulse}

We investigate the magnetization switching via a source-drain current pulse \cite{Yasuda17}. Under a static gate voltage \cite{Fan16}, we apply a step-like voltage pulse of width $t_{\rm S}$ to the source electrode, as shown in Fig.~\ref{fig:VS}(a) (see the upper panel). The applied gate voltage can reduce the energy barrier for the magnetization reversal due to the VCMA effect in Eq.~(\ref{Kua}). The resulting time evolution of ${\bf m}(t)$ is shown in Fig.~\ref{fig:VS}(a) with the constant gate voltage turned on at $t = 0$. The lower panel shows clearly that $m_z$ changes its sign by the pulsed $V_{\rm S}$ inputs, demonstrating the out-of-plane magnetization switching. Furthermore, applying subsequent pulses switches the magnetization direction faithfully, and the change is independent of the pulse's sign. The estimated switching time between points 1 and 2 (3 and 4) is $\sim18$~ns. In this simulation with static $V_{\rm G} = 0.32$~V and $V_{\rm S} = 1.5$~V, the magnetic anisotropy field $B_K$ and effective field due to the DL(FL) SOT $B_{\rm DL(FL)} = \Delta\mu_{x(y)}/(M_{\rm s}d)$ are evaluated for $m_z = 1$ as $B_K = 21$~mT, $B_{\rm DL} = 6.0$~mT, and $B_{\rm FL} = 17$~mT, respectively. Note that the calculated magnetic anisotropy field is $B_K = 136$~mT at $V_{\rm G} = 0$. The estimated current density (corresponding to $B_{\rm DL(FL)}$) is $J_{\rm S} = 1.9\times10^5$~A/cm$^2$ (see Sec.~\ref{FET} for details), which is consistent with those of TI-based magnetic heterostructures \cite{Fan14, Yasuda17, Han17, Wang17, Dc18, Wu19}. Figure~\ref{fig:VS}(b) and (c) show the magnetization switching trajectories during the durations shown by numbers (1--4) in the left upper panel of Fig.~\ref{fig:VS}(a). An equilibrium magnetization almost along $\pm\hat{\bf z}$ is rotated steeply around the current-induced spin polarization (${\bm \mu}$) by the pulsed SOT until the black star ($\star$), after which $B_K$ gradually stabilizes the magnetization with oscillations around the easy axis.

\subsection{Switching via a pulsed gate voltage $V_{\rm G}$}\label{G pulse}

We also investigate the magnetization switching via a pulsed gate voltage. Under a constant source-drain bias, we apply a step-like voltage pulse of width $t_{\rm G}$ to the gate electrode, as shown in Fig.~\ref{fig:VG}(a) (see the upper panel). The source-drain bias induces the spin polarization that takes the role of a constant-bias magnetic field \cite{Shiota12}, while the pulsed gate voltage reduces the energy barrier for the magnetization reversal via the VCMA effect during its duration. The resulting time evolution is shown in Fig.~\ref{fig:VG}(a) with the source-drain bias turned on at $t = 0$. Figure~\ref{fig:VG}(b) and (c) show the corresponding magnetization switching trajectories in which an equilibrium magnetization at an initial state (1 or 3) is rotated steeply around the bias effective magnetic field (proportional to ${\bm \mu}$) by the SOT until the black star, after which $B_K$ ($=136$~mT at $V_{\rm G} = 0$) aligns the magnetization direction with the easy axis by the Gilbert damping. As shown, the $z$-component of the magnetization changes sign by the pulsed $V_{\rm G}$ inputs, demonstrating the out-of-plane magnetization switching. The estimated switching time from the initial to final states is almost the same as the pulse width $t_{\rm G}\sim6.7$~ns. In this simulation with static $V_{\rm G} = 0.38$~V and $V_{\rm S} = 0.3$~V, we evaluate $B_K = 3.0$~mT, $B_{\rm DL} = 1.2$~mT, and $B_{\rm FL} = 3.6$~mT for $m_z = 1$. We emphasize that the current density corresponding to the SOT is $J_{\rm S} = 4.1\times10^4$~A/cm$^2$, which is smaller than that of the pulsed SOT method discussed above and those of TI-based magnetic heterostructures reported to date \cite{Wu19}.

\subsection{Switching phase diagram}\label{phase diagram}

\begin{figure}[ptb]
\begin{centering}
\includegraphics[width=0.55\textwidth,angle=0]{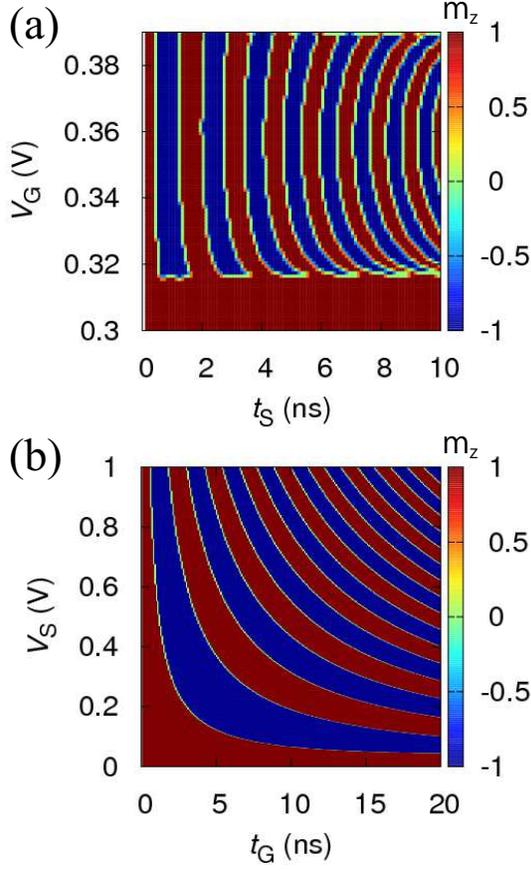} 
\par\end{centering}
\caption{
(a) Final-state diagram of $m_z$ at $V_{\rm S} = 1.5$~V as function of pulse duration time $t_{\rm S}$ and gate voltage $V_{\rm G}$. 
(b) Final-state diagram of $m_z$ at  $V_{\rm G} = 0.39$~V as function of pulse duration time $t_{\rm G}$ and source-drain voltage $V_{\rm S}$. 
}
\label{fig:Pswitch}
\end{figure}

We conclude this study with a guide for realizing the proposed magnetization switching methods. In particular, experimenters may be interested in how the final-state solution of $m_z$ depends on the input pulse width and voltages. In the following plots, we use the parameters for Cr-doped BST.
Figure~\ref{fig:Pswitch}~(a) shows the phase diagram of $m_z$ for a source-drain current pulse as a function of both $t_{\rm S}$ and $V_{\rm G}$. The diagram is calculated up to $V_{\rm G} \approx 0.39$~V to which the Fermi level reaches the bottom of a bulk conduction band. 
The final-state solution of $m_z$ oscillates rapidly depending on $t_{\rm S}$ rather than $V_{\rm G}$, whereas the diagram has a threshold $V_{\rm G}$ at the vicinity of 0.32~V at which the SOT competes with the anisotropy field. 
In Fig.~\ref{fig:Pswitch}~(b), we also show the phase diagram of $m_z$ for a pulsed gate voltage as a function of both $t_{\rm G}$ and $V_{\rm S}$. Clearly, the final-state solution of $m_z$ oscillates depending on both $t_{\rm G}$ and $V_{\rm S}$, whereas switching tends to succeed in the short pulse region of sub-nanosecond order. Consequently, switching will be achieved in the wide pulse duration between the nano- and sub-microsecond scales.
In practice, the proposed device would be mounted by combination with semiconductor devices such as CMOS whereas the switching speed of VLSI (very large-scale integration) is not so fast at present. Hence, control with a pulse width of a few nano-second is considered realistic. Figure~\ref{fig:Pswitch}~(b) shows that the controllability of the $t_{\rm G}$ pulse is better than the $t_{\rm S}$ pulse because of the width of the pulse. From the viewpoint of the speed, it can be expected that the $t_{\rm G}$ pulse has better compatibility with VLSI.

\section{Discussion}

\subsection{Source-drain current and Hall voltage: FET and memory operations}\label{FET}

\begin{figure}[ptb]
\begin{centering}
\includegraphics[width=0.47\textwidth,angle=0]{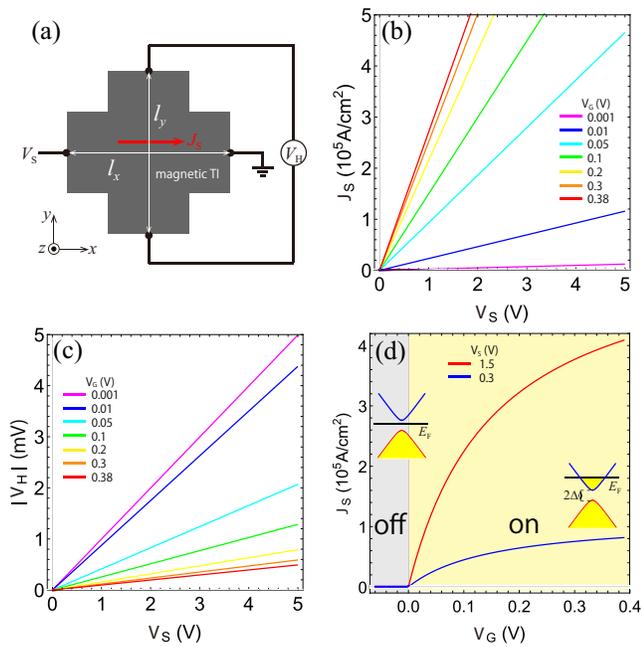} 
\par\end{centering}
\caption{
(a) Top view of device proposed in Fig.~\ref{fig:device} (here the gate and dielectric layers are hidden), where $l_x$ and $l_y$ are the lengths of the source-drain and Hall directions, respectively. (b) Source-drain current density $J_{\rm S}$ of proposed device versus $V_{\rm S}$ for different values of $V_{\rm G}$. (c) Corresponding Hall voltage $|V_{\rm H}|$ versus $V_{\rm S}$ for different values of $V_{\rm G}$. (d) Transfer characteristics obtained for different $V_{\rm S}$, indicating the FET operation (on/off) of the proposed device. Insets show the band pictures corresponding to the on/off states. In these plots, we assume $|m_z| = 1$, $d = 7$~nm, and $l_x = l_y = 10$~$\mu$m.
}
\label{fig:JSVH}
\end{figure}

To evaluate the magnitude of current density ($J_{\rm S}$) realizing the magnetization switching in Sec.~\ref{switching}, we calculate the source-drain current flowing on the TI surface in Fig.~\ref{fig:JSVH}(a). We also calculate the magnitude of the output Hall voltage ($V_{\rm H}$) to read out a direction of the out-of-plane magnetization by its sign \cite{Fujita11}. According to Eq.~(\ref{sigma}) and $E_x = V_{\rm S}/l_x$, the corresponding quantities can be written as 
\begin{align}
J_{\rm S} & = \frac{\sigma_{\rm L}(V_{\rm G})}{d}\frac{V_{\rm S}}{l_x}
, \label{JS}\\
V_{\rm H} & = \frac{\sigma_{\rm AH}(V_{\rm G})}{d}\frac{V_{\rm S}}{l_x}l_y
, \label{VH}
\end{align}
which are plotted in Fig.~\ref{fig:JSVH}(b) and (c), respectively. In these plots, we use the same parameters as for Cr-doped BST in Sec.~\ref{switching}. As a reminder, note again that $n_{\rm int} \approx 0$~cm$^{-2}$ (corresponding to $E_{\rm F}(V_{\rm G} = 0) = \Delta = 30$~meV) is assumed for the electrostatic equilibrium. Figure~\ref{fig:JSVH}(d) shows the FET operation (on/off) of the proposed device. Clearly, we can switch the source-drain current by a reasonable gate voltage compared with modern FET devices. Therefore, combining this FET operation with the proposed magnetization-switching method promises an FET with the functionality of a nonvolatile magnetic memory \cite{Takiguchi19}. The bit stored in this device is read out by measuring $V_{\rm H}$ and determining its sign. So, if $E_{\rm F}(V_{\rm G} = 0)$ is tuned within the surface band gap by the element substitution \cite{Yoshimi15}, the quantum anomalous Hall effect might allow the readout process to be free from energy dissipation due to Joule heating.

\subsection{TI/FI bilayers}\label{TI/FI}

We discuss the possibility of magnetization switching in TI/FI bilayers with the magnetic proximity effect at the interface \cite{Jiang15, Hirahara17, Fanchiang18}. A device corresponding to Fig.~\ref{fig:device} is proposed by replacing the dielectric with an FI and the TI/magnetic-TI bilayer with a TI. In this case, it is easily shown that an exchange interaction between interface Dirac electrons and localized moments of the FI appears in the same form as that of the Dirac PMA given by Eq.~(\ref{Kua}) \cite{Semenov12, Tserkovnyak15}. Then, the magnetization dynamics can be analyzed by using the LLG equation (\ref{LLG}) involving crystalline magnetic anisotropies ($K_{\rm CMA}$) of the FI. Switching methods similar to those discussed in Sec.~\ref{switching} will be achieved for a magnetically almost-isotropic FI with small net magnetization (desirably $K_{\rm CMA}d/K_u \ll 1$ and $M_{\rm s} \lesssim 10^{5}$~Am$^{-1}$), reducing the demagnetizing field effect. The potential candidates for the FI layer are 2D van~der~Waals ferromagnetic semiconductors \cite{Khan19, Mogi19} and rare-earth iron garnets \cite{Tang16, Yang19, Leon18} near their compensation point, where the intrinsic magnetic anisotropy and magnetostatic field become inferior to those of the Dirac PMA. However, the compensation points of the candidates are at a finite temperature, therefore the thermal excitation of the TI bulk state and the effective magnetic field due to the thermal fluctuation of localized moments might affect the switching probability, which is beyond the scope of the present investigation. An alternative might be to use an insulating antiferromagnet with the A-type layered structure \cite{Li17} that has no magnetostatic field because of the tiny net magnetization. Note that a recent experiment reports that the TI surface states on ${\rm Bi_2Se_3}$ produce a PMA in the attached soft ferrimagnet ${\rm Y_3Fe_5O_{12}}$ \cite{Liu20}.

Herebefore, we have focused only on the all-insulating systems (i.e., TI and FI), whereas  TI/ferromagnetic-metal(FM) bilayer systems are important for spintronic applications and experiments. It is necessary to pay attentions for the direct application of our model to the TI/FM bilayer represented by a pair of ${\rm Bi_2Se_3}$ and Py because of the following two reasons.
First, our 2D model cannot capture the three-dimensional (3D) nature of the transport in the TI/FM bilayer. Indeed, in ${\rm Bi_2Se_3}$/Py bilayers, most of the electric current shunt through the Py layer and conductive bulk states of the ${\rm Bi_2Se_3}$ layer, which reduces the portion of the current interacting with the TI interface state. From this viewpoint, {\it Fischer}~{\it et~al.}\ \cite{Fischer16} show that in the FM layer spin-diffusion transport perpendicular to the interface plays a crucial role to generate the DL torque. In contrast, based on a 3D tight-binding model of the TI/FM bilayer, {\it Ghosh}~{\it et~al.}\ \cite{Ghosh18} demonstrate that a large DL SOT is generated by the Berry curvature of the TI interface state rather than the spin Hall effect of the bulk states.
Secondly, orbital hybridization between the 3$d$ transition metal and TI deforms the TI surface states, which shifts the Dirac point to the lower energy and generates Rashba-like metallic bands across $E_{\rm F}$ \cite{Zhang16,Marmolejo17}. Besides, the hexagonal warping effect might be important for ${\rm Bi_2Se_3}$ with a relatively large $E_{\rm F}$ due to its crystal symmetry. According to {\it Li}~{\it et~al.}\ \cite{Li19}, the Berry curvature for hexagonal warping bands involves not only out-of-plane magnetization components but also those of the in-plane, which implies that the in-plane magnetization can contribute to the DL SOT. Note that the hexagonal warping term is important under threefold-rotational symmetry as the ${\rm Bi_2Se_3}$ crystal structure while it becomes small in bulk insulating TIs (our focus) such as BSTS and Cr-doped BST due to reduction of the symmetry by the elemental substitution \cite{Arakane12}.

\subsection{Influence of finite temperatures}\label{temperature}

For the experimental probe of our proposal, one may be interested in how finite temperatures affect the magnetization switching. In our model, there are mainly three temperature effects: (i) temperature ($T$)--dependence of physical quantities of TIs ($\sigma_{\rm L(AH)}$ and $K_u$), (ii) the thermal excitation of the TI bulk states at finite temperatures, and (iii) a random magnetic field due to the thermal fluctuation of localized moments, potentially leading to a switching error. The cases (i) and (ii) attribute to electronic properties while the case (iii) is in usual treated by magnetization dynamics.

\begin{figure}[ptb]
\begin{centering}
\includegraphics[width=0.57\textwidth,angle=0]{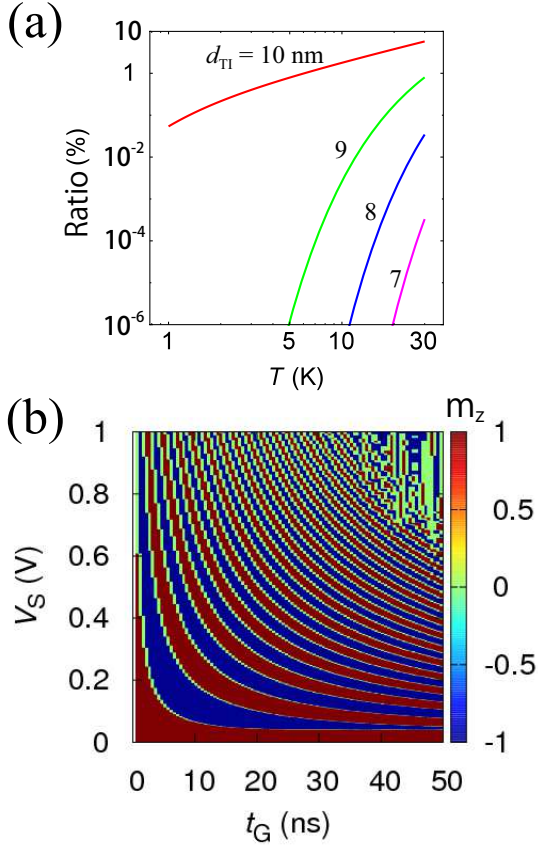} 
\par\end{centering}
\caption{
(a) Ratio~=~$n_{\rm b}d_{\rm TI}/n_{\rm s}$ as a function of $T$ for different magnetic TI thickness $d_{\rm TI}$. 
(b) Final-state diagram of at $T = 20$~K as function of pulse duration time $t_{\rm G}$ with  $V_{\rm G} = 0.39$~V and source-drain voltage $V_{\rm S}$. The influence of the thermal fluctuation emerges in a longer $t_{\rm G}$ region with lager $V_{\rm S}$. 
}
\label{fig:Pswitchth}
\end{figure}

Regarding the case (i), at low temperatures that satisfies $k_{\rm B}T \ll \Delta < E_{\rm F}$~($k_{\rm B}$ is the Boltzmann constant), we can regard the Fermi--Dirac distribution function $f(E) \approx \Theta(E - E_{\rm F})$ as the step function and then ignore $T$--dependence of $\sigma_{\rm L(AH)}$ and $K_u$. Indeed, when we set $\Delta = 30$~meV and $30~{\rm meV} \leq E_{\rm F} \leq 100$~meV for the system, the above condition is satisfied below 30~K~($k_{\rm B}T \approx 2.6$~meV) that is a temperature less than $T_{\rm c}$ of magnetic TIs or FIs discussed in Sec.~\ref{TI/FI}. 
Regarding the case (ii), based on the model of Ref.~\onlinecite{Chiba19}, we investigate the contribution of the bulk states in terms of the electron density by using 
\begin{align}
{\rm Ratio} = \frac{n_{\rm b}d_{\rm TI}}{n_{\rm s}}
, \label{Ratio}
\end{align}
where $n_{\rm b}$ is the bulk electron density of the TI with thickness $d_{\rm TI}$ and $n_{\rm s}$ is the surface electron density. Here, we assume that the bulk state of a TI thin film is a quantum well system along the $z$ direction, namely $E_{ks,\ell} =  \hbar^2/(2m^*)\left(  k_x^2 + k_y^2\right) + E_{z,\ell}(d_{\rm TI}) + \Delta_{\rm c} + s\Delta_{\rm b}$ with $E_{z,\ell}(d_{\rm TI}) = \hbar^2/(2m^*)\ell^2\pi^2/d^2$,
where $\ell$ is an integer, $m^*$ is the effective electron mass, and $\Delta_{\rm b}$ is the exchange energy in the bulk of magnetic TIs with spin index $s=\pm$. The corresponding electron density is given by $n_{\rm b/s} = \int_{-\infty}^\infty dED_{\rm b/s}(E)f(E)$, where $D_{\rm b/s}(E) = \sum_{ks}\delta\left(E-E_{\rm b/s}\right)$ is the bulk ($E_{\rm b} = E_{ks,\ell}$)/surface ($E_{\rm s} = E_{{\bf k}s}$ in Eq.~(\ref{E2d})) density of states. The computed result is shown in Fig.~\ref{fig:Pswitchth}~(a). Consequently, we find that Eq.~(\ref{Ratio}) is less than 10~\% below 30~K for a Cr-doped BST thin film with $E_{\rm F} = 95$~meV (corresponding carrier density $\sim10^{12}$~cm$^{-2}$), $\Delta_{\rm b} = 30$~meV, and $m^*/m_0 = 0.15$~($m_0$ is the free-electron mass) \cite{Butch10}, which implies that the thermal excitation of the TI bulk states is negligible below 30~K.

Regarding the case (iii), we perform similar simulations including the random magnetic field at $T = 20$~K with an isotropic 3D Gaussian distribution \cite{Shiota12}
\begin{align}
h_{\rm th} & = \sqrt{\frac{2k_{\rm B}T\alpha_{\rm eff}}{\mathcal{V}M_{\rm s}\gamma\left(  1 + \alpha_{\rm eff}^2\right)\varDelta t}}
, \label{hth}
\end{align}
where $\mathcal{V} \approx l_xl_yd = 10^{-18}$~m$^3$ is the volume of the magnet, $\varDelta t = 0.1$~ps, and parameters for Cr-doped BST are used. First, we investigate the effect of Eq.~(\ref{hth}) on the pulsed $J_{\rm S}$ induced magnetization switching and find that the random magnetic field does not affect the switching diagram of Fig.~\ref{fig:Pswitch}~(a) up to 20 K. We also test the case of the pulsed $V_{\rm G}$ driven magnetization switching and show the result in Fig.~\ref{fig:Pswitchth}~(b) for the extended input parameter ranges. As seen, the influence of Eq.~(\ref{hth}) emerges in the longer pulse region with lager $V_{\rm S}$, reflecting randomness of the thermal fluctuation, whereas we confirm that stable magnetization switching is achieved in the wide parameter range. Note that the thermal stability of magnet $K_u\mathcal{V}/(k_{\rm B}T) \approx 5.2\times10^5$ at $E_{\rm F} = 30$~meV ($V_{\rm G} = 0$) and $T = 20$~K satisfies the required condition ($> 60$) for a non-volatile memory.

\section{Summary}

In summary, we have presented two distinct methods for magnetization switching by using electric-field control of the SOT and MAE in TI/magnetic-TI hybrid systems. We formulated analytically the uniaxial magnetic anisotropy in magnetic TIs as a function of the Fermi energy and showed that the out-of-plane magnetization is always favored for the partially occupied surface band. We further proposed a transistor-like device with the functionality of a nonvolatile magnetic memory adopting (i) the VCMA writing method that requires no external magnetic field and (ii) read-out based on the anomalous Hall effect. For the magnetization reversal, by using parameters of Cr-doped BST, the estimated source-drain current density and gate voltage were of the orders of $10^4$--$10^5$~A/cm$^2$ and $0.1$~V, respectively, below 20~K. As a conclusion of this study, we showed the switching phase diagram for the input pulse width and voltages as a guide for realizing the proposed magnetization-reversal method. We also discussed the possibility of magnetization switching by the proposed method in TI/FI bilayers with the magnetic proximity effect. Similar magnetization switching may be achieved by the FI layer with 2D van~der~Waals ferromagnetic semiconductors or rare-earth iron garnets near their compensation point. However, the compensation points of the FIs are at a finite temperature, so the thermal excitation of the TI bulk state and the effective magnetic field due to the thermal fluctuation of localized moments might affect the switching probability, which is beyond the scope of the present investigation. Simultaneous control of the magnetic anisotropy and SOT by an electric gate may lead to low-power memory and logic devices utilizing TIs.

\section{Acknowledgments}

The authors thank Yohei Kota, Koji Kobayashi, Seiji Mitani, Jun'ichi Ieda, and Alejandro O. Leon for valuable discussions.
This work was supported by Grants-in-Aid for Scientific Research (Grant No.~20K15163  and No.~20H02196) from the JSPS.

\appendix

\section{Current-expression of SOT}\label{CESOT}

We rewrite the SOT in terms of a current density flowing on the magnetized TI surface. According to Eq.~(\ref{spin2}) in the main text, the current-induced spin polarization involving ${\bf J}_{\rm S} = \sigma_{\rm L}{\bf E}/d = (\sigma_{\rm L}/d)(V_{\rm S}/l_x)\hat{\bf x}$ (in units of Am$^{-2}$) is given by
\begin{align}
{\bm \mu} = -\frac{d}{ev_{\rm F}}\left(  -\theta_{\rm AH}{\bf J}_{\rm S} + \hat{\bf z}\times{\bf J}_{\rm S}\right)
,\label{spin3}
\end{align}
where 
\begin{align}
\theta_{\rm AH} 
& \equiv \frac{\sigma_{\rm AH}}{\sigma_{\rm L}} = \frac{8\hbar}{E_{\rm F}\tau}\frac{\Delta m_zE_{\rm F}\left(  E_{\rm F}^2+\Delta^2m_{z}^2\right)}{\left(  E_{\rm F}^2-\Delta^2m_{z}^2\right)\left(  E_{\rm F}^2+3\Delta^2m_{z}^2\right)}
\label{AHangle}
\end{align}
is an anomalous Hall angle.
Inserting Eq.~(\ref{spin3}) into Eq.~(\ref{SOT}) in the main text, we obstinate the current-expression of SOT
\begin{align}
{\bf T}_{\rm SO} = \frac{\gamma\Delta}{ev_{\rm F}M_{\rm s}}{\bf m}\times\theta_{\rm AH}{\bf J}_{\rm S} - \frac{\gamma\Delta}{ev_{\rm F}M_{\rm s}}{\bf m}\times\left(  \hat{\bf z}\times{\bf J}_{\rm S}\right)
.\label{SOT3}
\end{align}
Recalling that the first term is responsible for the DL-torque associated with a magnetoelectric coupling, one may expect that a giant current-induced SOT is obtained by Eq.~(\ref{AHangle}) in the case of $E_{\rm F} \approx \Delta m_z$. However, ${\bf J}_{\rm S}$ then becomes nearly zero because the relation between $\theta_{\rm AH}$ and ${\bf J}_{\rm S}$ is tradeoff [see Eq.~(\ref{sigma}) in the main text], and therefore in Eq.~(\ref{AHangle}) one should not seek the reason of a giant current-induced SOT in recent experiments \cite{Fan14,Mellnik14,HYang15}.

\section{Electric-field effect on $E_{\rm F}$}\label{EFEonEF}

We model the electrical modulation of $E_{\rm F}$ by the gate voltage $V_{\rm G}$ of the device proposed in the main text (Fig.~\ref{fig:device}). According to Gauss's law, an induced charge accumulation per unit area ($q_e$) on the TI surface is described by 
\begin{align}
\frac{q_e}{\epsilon} = \frac{V_{\rm G}}{d_{\rm D}}, \label{qe} 
\end{align}
where $q_e = en_c$ with the modulated electron density $n_c = \left( E_{\rm F}(V_{\rm G})^2 - E_{\rm F}(0)^2\right)/4\pi\left( \hbar v_{\rm F}\right)^2$ \cite{Chiba19APL}. Defining the intrinsic electron density as $n_{\rm int} = \left(  E_{\rm F}(0)^2 - \Delta^2\right)/4\pi\left( \hbar v_{\rm F}\right)^2$ and inserting it into Eq.~(\ref{qe}), we finally obtain Eq.~(\ref{EVG}) in the main text describing the electric-field effect on $E_{\rm F}$.


\end{document}